\documentclass{ws-ijmpa}

\begin{document}

\markboth{Marta Tichoruk}
{Pion and kaon production in nucleon - nucleon collisions}

%
\catchline{}{}{}{}{}
%

\title{PION AND KAON PRODUCTION \\ IN NUCLEON - NUCLEON COLLISIONS.}

\author{MARTA TICHORUK }

\address{Institute of Physics, Jagiellonian University \\
PL-30-059 Cracow, Poland} 

\author{ANTONI SZCZUREK}

\address{Institute of Nuclear Physics\\
PL-31-342 Cracow, Poland\\
and\\
University of Rzesz\'ow\\
PL-35-959 Rzesz\'ow, Poland\\
Antoni.Szczurek@ifj.edu.pl}

\maketitle

\begin{history}
\received{10 August 2006 }
\revised{Day Month Year}
\end{history}

\begin{abstract}
Inclusive cross section for pion production in proton - proton
collisions are calculated based on unintegrated parton distribution
functions (uPDFs). In addition to purely gluonic terms the present 
approach includes also quark degrees of freedom.
Phenomenological fragmentation functions from the literature are used. 
The new mechanisms are responsible for $\pi^+$ - $\pi^-$ asymmetry. 
In contrast to standard collinear approach, application of
2 $\to$ 1 $k_t$ - factorization approach can be extended towards
much lower transverse momenta, both at mid and forward rapidity
region.
The results of the calculation are compared with SPS and RHIC data. 
\keywords{unintegrated parton distributions; fragmentation functions; inclusive cross section.}
\end{abstract}

\ccode{PACS numbers: 11.25.Hf, 123.1K}

\section{Introduction}	
The distributions of mesons at large transverse momenta
in $p p$ or $p \bar p$ collisions are usually calculated
in the framework of perturbative QCD using collinear factorization
(see e.g.\cite{Owens,Field,EK97,EH02}). While the shape at
transverse momenta larger than 2-4 GeV can be relatively
well explained, there are discrepancies at lower transverse momenta.
In this analysis, the calculations are performed using a new approach,
based on the unintegrated parton distributions.
In recent years only  gluon degrees of freedom are taken explicitly
in this context \cite{KL01}.   
In the present analysis, in addition to the $gg \to g$ mechanism, 
We include also $q_f g \to q_f$ and $g q_f \to q_f$ mechanisms
and similar ones for antiquarks, in order to obtain a fully
consistent description. The new contributions $q_f g \to q_f$ and
$g q_f \to q_f$ are comparable to the contribution of the $gg \to g$
diagram at midrapidities and are dominant in the
fragmentation region. The new mechanisms are responisible for
$\pi^+ - \pi^-$ asymmetry. A purely gluonic mechanism would lead
to identical production
of positively and negatively charged hadrons.
The recent results of the BRAHMS experiment
\cite{BRAHMS} show that the $\pi^-/\pi^+$
and $K^-/K^+$ ratios differ from unity. This put into
question the successful description of Ref.\cite{KL01}.
In the light of this experiment, it becomes obvious that
the large rapidity regions have more complicated flavour structure.
At lower energies these ratios are known to differ from unity
drastically \cite{Antreasyan}.
Many unintegrated gluon distributions in the literature
are ad hoc parametrizations of different sets of
experimental data rather than derived from QCD.
Recently Kwieci\'nski and collaborators
\cite{CCFM_b1,CCFM_b2,GKB03} have shown how to solve the so-called
CCFM equations by introducing unintegrated parton distributions in
the space conjugated to the transverse momenta \cite{CCFM_b1}.
We present results for pion and kaon production based on the unintegrated
parton (gluon, quark, antiquark) distributions.
\section{Inclusive cross section for partons}
The formulae for inclusive quark/antiquark
distributions are similar to the formula for $gg \to g$ \cite{GLR81}

%
\begin{eqnarray}
&&\frac{d \sigma^{A}}{dy d^2 p_t} = \frac{16  N_c}{N_c^2 - 1}
{\frac{1}{p_t^2}}
 \nonumber \\
&& \int
 \alpha_s({\Omega^2}) \;
{  f_{g/1}(x_1,\kappa_{1_t}^2,\mu^2)} \;
{  f_{g/2}(x_2,\kappa_{2_t}^2,\mu^2)}
\nonumber \\
&&\delta^{(2)}(\vec{\kappa}_{1_t}+\vec{\kappa}_{2_t} - \vec{p}_t)
\; d^2 \kappa_{1_t} d^2 \kappa_{2_t}    \; ,
\label{diagram_A_tr}
\end{eqnarray}
These seemingly 4-dimensional integrals can be written
as 2-dimensional integrals after a suitable change
of variables \cite{szczurek03}. 
The formulae can be written in the equivalent way
in terms of the parton distributions in the space conjugated
to the transverse momentum \cite{CS05}.
\section{Inclusive cross section for hadrons}
There are a few sets of fragmentation functions available in
the literature, details were described e.g. in \cite{BKK95},
\cite{Kretzer2000}, \cite{AKK05}.
The inclusive distributions of hadrons (pions, kaons, etc.)
can be obtained through a convolution of inclusive distributions
of partons and flavour-dependent fragmentation functions.
One dimensional distributions of hadrons can be obtained
through the integration over the other variable (see \cite{CS05}).
\section{Results}
\begin{figure}[htb]
\epsfig{file=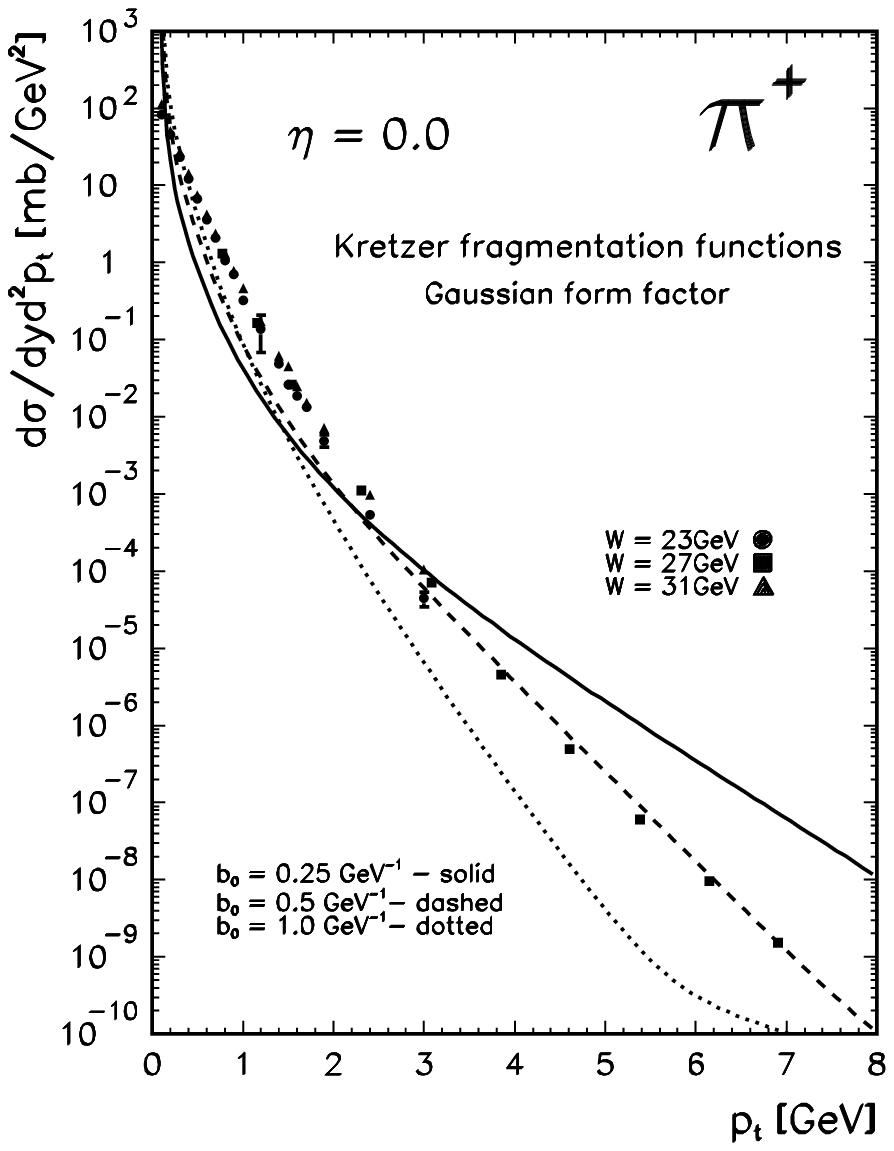,width=3.8cm}
\epsfig{file=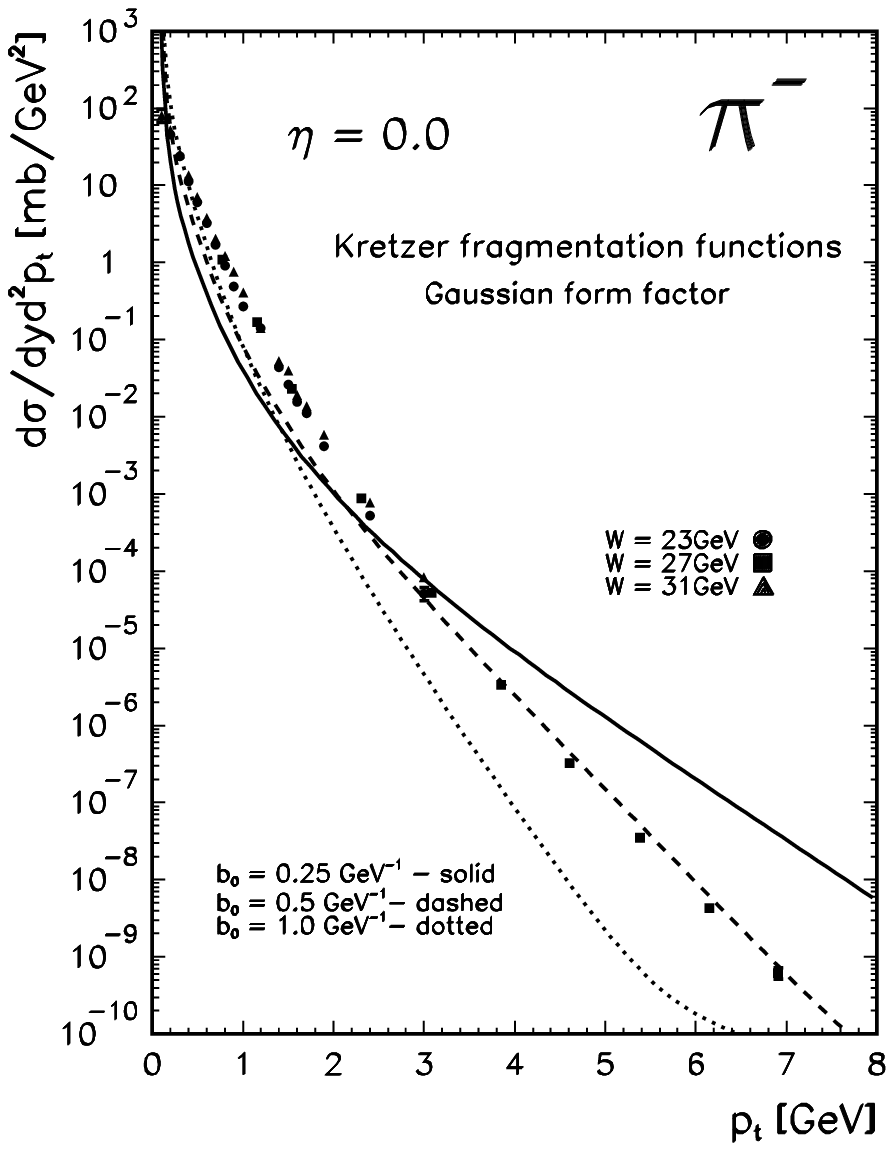,width=3.8cm}
\epsfig{file=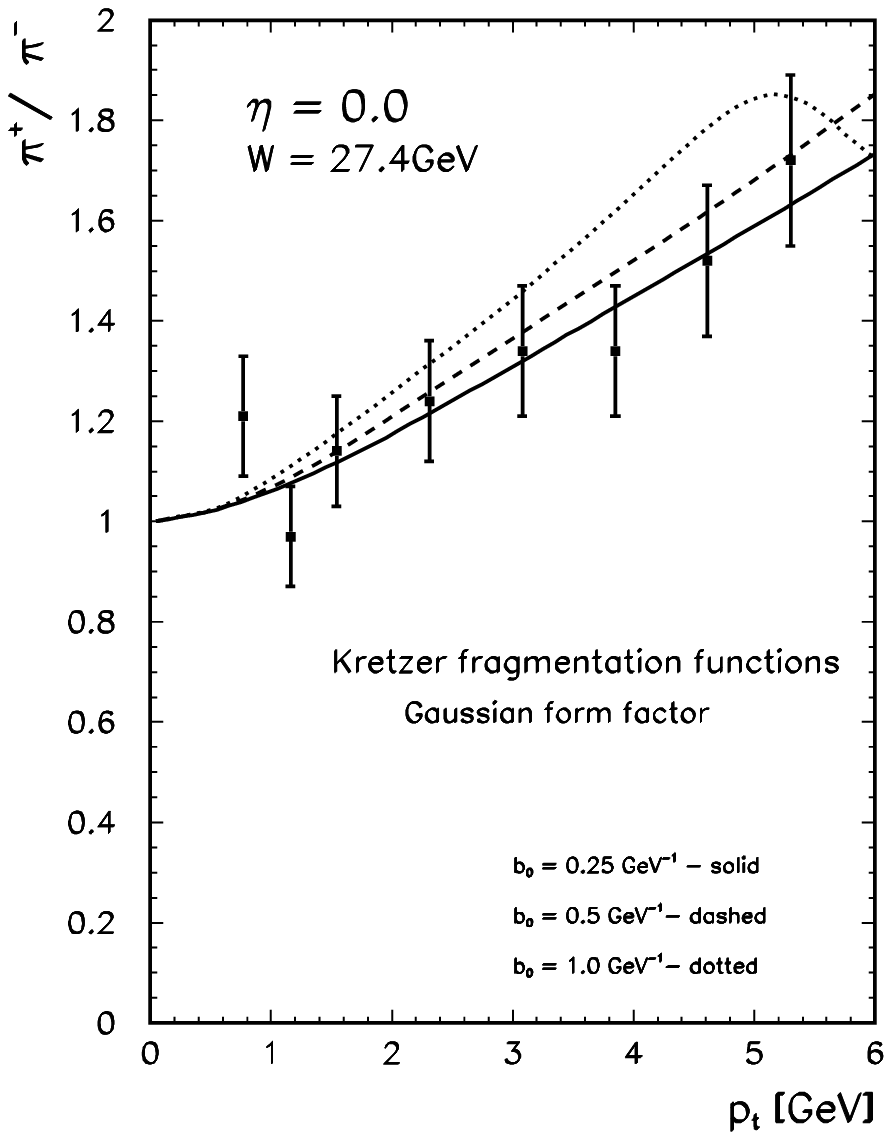,width=3.8cm}
\caption[*]{
Invariant cross section as a function of transverse momentum
of $\pi^+$ (left panel) and $\pi^-$ (right panel) for $\eta$ = 0,
W = 27.4 GeV and different values of parameter $b_0$ of the Gaussian
form factor.
Experimental data for W  = 23, 31 GeV from \cite{Alper} and
for W = 27.4 GeV \cite{Antreasyan} are shown for comparison. 
\label{fig:pions_pt}
}
\end{figure}

In Fig.~\ref{fig:pions_pt} we compare the model invariant cross
sections for $p p \to \pi^+$ (left panel) and $p p \to \pi^-$
(middle panel) as a function of pion transverse momentum
at W = 27.4 GeV for different values of the parameter $b_0$
of our Gaussian nonperturbative form factor (for explanation 
see \cite{CS05}) with
the experimental data from Ref.\cite{Alper}, \cite{Antreasyan}.
In principle, the result should not exceed experimental data
especially in the perturbative regime of $p_t >$ 2 GeV where
the perturbative $2 \to 2$ parton subprocesses are crucial.
This limits the value of the nonperturbative
form factor to $b_0 >$ 0.5 GeV$^{-1}$.
Inclusion of diagrams $B_1$ (q$_f$ g $\to$ q$_f$)
and $B_2$ (g q$_f$ $\to$ q$_f$)
(see \cite{CS05}) in conjunction
with the flavour dependent fragmentation functions leads to
the $\pi^+ - \pi^-$ asymmetry. In the right panel
of Fig.\ref{fig:pions_pt}
we show  the asymmetry as the function of pion transverse momentum.
The asymmetry is well described by this model, in contrast to
individual distributions. This seems to suggest the right relative
contributions of diagram $A$ (gg $\to$ g), $B_1$ and $B_2$.
The asymmetry depends only weakly on the value of the parameter
$b_0$ of the Gaussian nonperturbative form factor.

\begin{figure}[htb]
  \centerline{\psfig{file=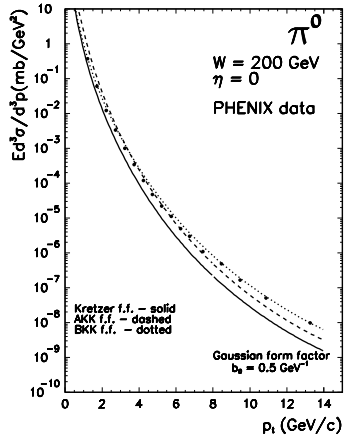,width=5.0cm}}
\caption[*]{
Invariant cross section for $\pi^0$ production as a function
of pion transverse momentum at W = 200 GeV and $\eta$ = 0.0.
The $k_t$-factorization results are shown
for different sets of fragmentation functions.
The experimental data of the PHENIX collaboration are
from \cite{PHENIX_pi0}. 
\label{fig:PHENIX_FF}
}
\end{figure}

The PHENIX collaboration has measured invariant cross section
as a function of the $\pi^0$ transverse momentum at W = 200 GeV
in a very narrow interval of pseudorapidity $\eta$ = 0.0 $\pm$ 0.15.
In Fig.\ref{fig:PHENIX_FF} we show full result
(diagrams $A$, $B_1$ and $B_2$ \cite{CS05},\cite{csrhic})
for different fragmentation functions
\cite{BKK95,Kretzer2000,AKK05}.

\begin{figure}[pb]
\begin{center}
\epsfig{file=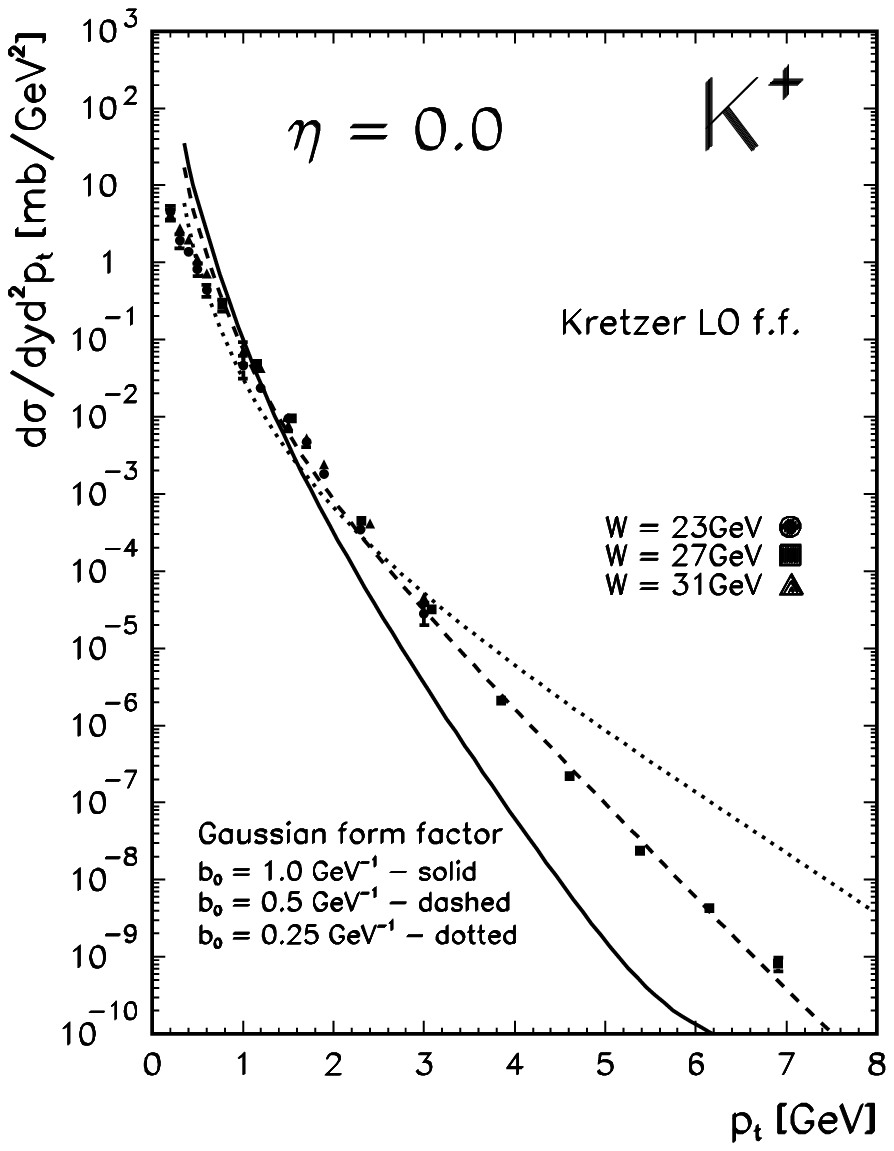,width=4.0cm}
\epsfig{file=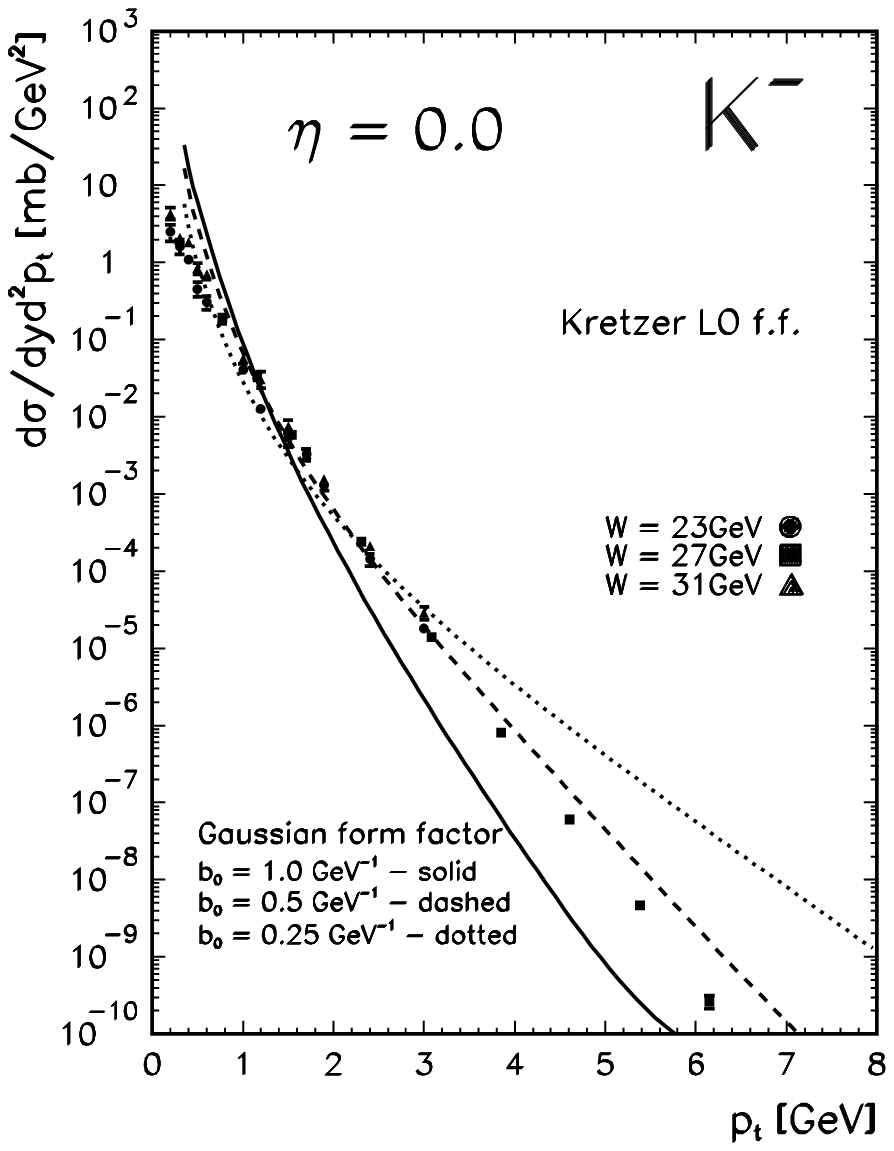,width=4.0cm}
\end{center}
\vspace*{8pt}
\caption{Dependence of the invariant cross section for kaon production
on the value of the parameter $b_0$ of the Gaussian nonperturbative
form factor.
\label{chk1}}
\end{figure}

In Fig.\ref{chk1} we present dependence of the invariant cross
section for kaon production,
calculated for W = 27.4 GeV,  on the value of the parameter $b_0$ of
the Gaussian nonperturbative form factor. The data for 
$p_t$ $>$ 0.5 GeV are well described by the $k_t$-factorisation
approach with the Kwieci\'nski UPDFs  
for the parameter $b_0$ $=$ 0.5 GeV$^{-1}$. 
This is the same value of the paramater as that obtained for pions
in the same energy range (see Fig. \ref{fig:pions_pt}).  
More examples can be found in \cite{csrhic} and \cite{doc_diss}.
\section{Conclusions}
The formalism based on uPDFs, which fulfill so-called  Kwieci\'nski evolution equations, provides a reasonable 
description of the experimental data, including SPS as well as
recent data of the PHENIX, BRAHMS and STAR collaborations.
A good agreement with experimental data is 
obtained, especially at relatively small transverse momenta and
large values of pseudorapidity.   
The mechanisms, which invole quark/antiquark degrees of freedom are
significant and lead to an asymmetry in 
the production of $\pi^+$ and $\pi^-$ mesons. \\\textmd{}

{\bf Acknowledgments}

This work was supported by the Polish State Committee for Scientific
Research under grant no. 1 P03B 097 29.

\end{document}